\documentstyle[preprint,aps]{revtex}
\begin{document}
\draft
\widetext

\title{
Current fluctuations in a single tunnel junction}
\author{Hyunwoo Lee\cite{HLee}}
\address{
Department of Physics, Massachusetts Institute of Technology,
  77 Massachusetts Ave., Cambridge, MA 02139}
\author{L. S. Levitov\cite{LSLevitov}}
\address{
Department of Physics, Massachusetts Institute of Technology,
  77 Massachusetts Ave., Cambridge, MA 02139,\\
and Landau Institute for Theoretical Physics,
     2 Kosygin St., Moscow 117334, Russia
}
\maketitle

\widetext
\begin{abstract}
We study noise spectra of currents through a tunnel junction
in weak tunneling limit. We introduce effective capacitance to
take into account the interaction effect and explicitly incorporate
the electromagnetic environment of the junction into the formulation.
We study the effect of charging energy and macroscopic environment
on noise spectra.
We calculate current fluctuations at tunneling barrier and
fluctuations measured at leads.
It is shown that two fluctuations have different noise spectra
and the relation between them is nontrivial.
We provide an explanation for the origin of the difference.
Experimental implications are discussed.
\end{abstract}


\narrowtext
\section{Introduction}
\label{sec:Introduction}
Even though ``noise'' usually represents unwanted
fluctuations which blur information,
noise is on the fundamental level closely
connected with the dynamics of a system
and it contains variety of information.

In this context, there have been many recent
studies on a noise spectrum of current in small devices.
In ballistic systems, the noise power can be suppressed
to zero due to Pauli exclusion
principle\cite{Lesovik,Yurke,Buttiker}.
In diffusive systems, the noise power is suppressed to
$1/3$ of the classical value\cite{Beenakker}.
There have been attempts to study the effect of electron-electron
interaction on the noise. For example,
noise spectrum in 1D systems was calculated from Luttinger liquid
approximation\cite{Chamon,Fendley}.

In this paper, we study current fluctuations of ultra-small
tunnel junctions made of two pieces of metal separated
by a thin insulating barrier.

It is helpful to discuss a few limiting cases.
It is well known that conventional tunnel junctions show
Ohmic behavior $I(V)=V/R_T$ when it is driven by
external voltage bias and the current noise spectrum
of Ohmic junctions follows Johnson-Nyquist formula\cite{Dahm}
\begin{equation}
S_{I_T}(\omega)={1 \over 2\pi R_T}\sum_{\pm}
  (\hbar \omega \pm eV)\coth\left({\hbar \omega \pm eV
  \over 2k_B T}\right).
\label{eqn:noiseDahm}
\end{equation}

Recent advances in fabrication technology have made
it possible to construct very small devices.
It was pointed that electron-electron
interaction effect becomes important for small devices.
For example, it was argued that
a tunnel junction develops tunneling gap in
current-voltage characteristic when
the temperature is lower than the charging energy of
a junction $E_C=e^2/2C$
\cite{Zorin,Averin,Geigenmuller,Likharev,Averin2}.

The charging energy was predicted to affect noise spectrum
as well. Ben-Jacob {\it et al}\cite{Ben-Jacob} studied
fluctuations of current in an open tunnel junction.
They calculated equilibrium
noise of a junction and found that
the low frequency noise $\hbar\omega < E_C$ is suppressed
to zero in the low temperature limit $k_B T \ll E_C$.

Following theoretical predictions on the charging effect
in a small tunnel junction, many experimental attempts
were made to observe the charging effect or so called
Coulomb blockade effect. Coulomb blockade effect was
verified for multi-junction systems
\cite{Lambe,Fulton,Bentum,Kuzmin,Wilkins,Delsing,Geerligs}.
However the clear verification of the effect
for a single tunnel junction has not been reported.

It was later realized that the electrical transport
properties of a single tunnel junction crucially
depends on the electromagnetic environment of a junction
which is decided by properties of leads
on junction chip\cite{Nazarov,Devoret,Girvin}.
Current-voltage characteristic was calculated
with explicit account of the environment
by modeling the environment as linear impedance
presented to the junction.
It was shown that the Coulomb blockade effect is erased
by the quantum fluctuations of charge
when the environment has low impedance
and that high impedance environment is essential
ingredient for the realization of Coulomb blockade effect.
The dependence on the electromagnetic environment was
experimentally observed\cite{Cleland}

Behaviors of single tunnel junctions are decided
not only by the ratio of temperature and the charging energy
but also by the environment.
It is a goal of this paper to extend the earlier works
to include noise spectrum calculation for general
environment.

More important goal of this paper is to show that noise
measurement in a single tunnel junction is quite nontrivial.
For illustration, we calculate current fluctuations at
the tunneling barrier and fluctuations at the leads.

Most theoretical works on noise
calculated the current fluctuations at a specific
point of a circuit, usually at a potential barrier
right in the middle of a junction.
However as pointed out by Landauer {\it et al}\cite{Landauer},
conventional noise measurements have no relation with the electrons
crossing the potential barrier of the junction. The noise is
measured, for example, through the voltage across a series resistor
in a circuit.

Noise measured in experiments is not precisely the same
quantity as the noise treated in various theoretical calculations.
Noise measured in experiments
is complicated by the intrinsic noise
of noise measuring device.
To make comparison with theoretical calculations,
it is usually assumed that the intrinsic noise makes
linear contribution and that experiments can achieve
agreement with theoretical calculations by subtracting off
the intrinsic noise from the measured noise.
We will call this {\it linear superposition assumption}
\cite{Landauer,Liefrink}.

In this paper, we assume that leads, which is connected to
a junction and provides electromagnetic environment, can
be used as a measuring device and calculate current
fluctuations at leads. We also calculate current
fluctuations at tunneling barrier and compare
them with current fluctuations at leads to test
the linear superposition assumption.
To briefly mention a result
of the calculation, we find that the linear superposition
assumption holds only when
\begin{equation}
|Z(\omega)| \ll  {1 \over \left| i\omega C \right| },
\label{eqn:criterion}
\end{equation}
where $Z(\omega)$ is the impedance of the environment.
In low impedance environment, the assumption breaks
down due to finite relaxation time of environment.
In high impedance environment, noise becomes quite
nontrivial and spectral density at $\hbar\omega \approx
\pm E_C$ can be reduced by applying small voltage bias.
We will later provide the physical origin of this
novel phenomena.

In Sec.\ref{sec:Formulation}, we present a formulation
of a problem which incorporates the charging energy and
the environment. The formulation allows simple perturbation
calculation in the weak tunneling limit.
In Sec.\ref{sec:ITvsIR}, we introduce two current
operators. One corresponds to current at
potential barrier of a junction, which we call
{\it tunneling current}. The other corresponds to
current measured at the leads, which we
call {\it relaxed current}. Expectation values of
the two operators are evaluated for a few examples of
environment.
In Sec.\ref{sec:tunnelingnoise}, the noise spectrum of
the tunneling current is calculated. It shows agreement
with existing theories in various limiting cases.
In Sec.\ref{sec:relaxednoise}, the noise spectrum of
the relaxed current is calculated.
It is compared to the noise spectrum of the tunneling
current and the difference is identified. The difference
is attributed to the finite relaxation time of
environment and nontrivial zero point
fluctuations of environmental degree of freedom.

\section{Formulation}
\label{sec:Formulation}
Fig.\ref{fig:circuit}(a) presents a schematic diagram of
a system which we consider in the paper. A junction
is represented by a capacitor notation with an arrow
right beside it to emphasize tunneling. The junction is
connected to the electromagnetic environment and voltage source.
It is assumed that the environment is linear
so that it can be modeled as an impedance $Z(\omega)$.
The impedance is determined by properties of leads connected
to junction chip. For detailed discussion of environment
and the model system in Fig.\ref{fig:circuit}(a),
see Ref.\cite{Grabert}.

We use the same formulation as the one adopted in Ref.\cite{Devoret}.
The tunnel junction connected to the environment is described
by the following hamiltonian(for more details about
this hamiltonian, see \cite{Ingold}),
\begin{equation}
H=H_{qp}+H_{env}+H_{T} \, .
\end{equation}
The first term $H_{qp}$ describes quasi-particle degrees of freedom,
\begin{equation}
H_{qp}=\sum_{k\sigma}(\epsilon_k +eV)c_{k\sigma}^{\dagger}
  c_{k\sigma}+\sum_{q\sigma}\epsilon_q c_{q\sigma}^{\dagger}
  c_{q\sigma} \, ,
\end{equation}
where $k$ and $q$ label the quasi-particle states in the left
and the right electrodes, respectively
and $\sigma$ represents spin degree of freedom.

The second term $H_{env}$ describes environmental degrees of freedom
and the charging energy,
\begin{equation}
H_{env}={\tilde{Q}^2 \over 2C} + \sum_{n=1}^{N}
  \left[ {q_n^2 \over 2C_n} + \left({\hbar \over e}\right)^2
  {1 \over 2L_n}\left(\tilde{\varphi}-\varphi_n\right)^2 \right] \, ,
\end{equation}
where $\tilde{Q}$ is the charge on the capacitor plates of
the junction with the steady state contribution subtracted off
$\tilde{Q}=Q-CV$. $\tilde{\varphi}$ is the phase variable
across the junction, $\tilde{\varphi}(t)=(e/\hbar)
\int_{-\infty}^{t} \, dt' \tilde{Q}(t')/C$.
$\tilde{Q}$ and $\tilde{\varphi}$ are conjugate variables
describing macroscopic state of the junction and
they satisfy the quantization relation,
\begin{equation}
\left[ \tilde{\varphi}, \tilde{Q} \right] = ie \, .
\label{eqn:commutation}
\end{equation}
Phase $\tilde{\varphi}$ is coupled to a set of harmonic oscillators
following the Caldeira-Leggett description\cite{Caldeira}
of the environment.

The third term describes the tunneling between two plates of
the capacitor,
\begin{equation}
H_T=\sum_{kq\sigma}T_{kq}c_{q\sigma}^{\dagger}c_{k\sigma}
  e^{-i\tilde{\varphi}} + \mbox{H.c.} \, ,
\end{equation}
where $T_{kq}$ is the tunneling matrix element.
It is simple to verify
that $e^{-i\tilde{\varphi}}$ is a charge lowering operator,
\begin{equation}
e^{i\tilde{\varphi}}\tilde{Q}e^{-i\tilde{\varphi}}
=\tilde{Q}-e \, .
\end{equation}
Because $e^{\pm i\tilde{\varphi}}$ explicitly takes care of
the charge increase and decrease,
the quasi-particle degrees of freedom and
the environment degrees of freedom can be separated
so that quasi-particle operators commute with the charge
and phase operators.

\section{Tunneling current and Relaxed current}
\label{sec:ITvsIR}
To calculate noise spectra, we need to construct current operators.
The tunneling current operator can be easily identified
from the operator equation of motion,
\begin{eqnarray}
\hat{I}_T &=& {i \over \hbar}\left[H,e\sum_{q\sigma}
   c_{q\sigma}^{\dagger}c_{q\sigma}\right]
   \label{eqn:Tcurrent} \\
   &=& -{e \over \hbar}\sum_{kq\sigma}\left(
   iT_{kq}c_{q\sigma}^{\dagger}c_{k\sigma}e^{-i\tilde{\varphi}}
   +\mbox{H.c.} \right) \, . \nonumber
\end{eqnarray}
This operator represents the net number of electrons tunneled
from one plate of the capacitor to the other.
Experimentally, $\hat{I}_T$ corresponds to current measured
right at the tunneling barrier(Fig.\ref{fig:circuit}(b)).

Eq.(\ref{eqn:Tcurrent}) is not a unique choice as
a current operator and there is another
way of constructing a current operator,
\begin{eqnarray}
\hat{I}_R &=& {i \over \hbar}\left[H_{env},\tilde{Q}\right]
   \nonumber \\
   &=& -{\hbar \over e}\sum_{n} {1 \over L_n}
      (\tilde{\varphi}-\varphi_n) \, .
\end{eqnarray}
For definiteness, we will call $\hat{I}_T$
{\it tunneling current} and
$\hat{I}_R$ {\it relaxed current}. These currents can be related
in a simple way,
\begin{equation}
\hat{I}_R-\hat{I}_T={i \over \hbar}\left[H, \tilde{Q} \right]
  = {d\tilde{Q} \over dt} \, ,
\end{equation}
and from this relation,
we interpret $\hat{I}_R$ as a charge flow at
the environment. Fig.\ref{fig:circuit}(b) illustrates
the difference between $\hat{I}_T$ and $\hat{I}_R$.
The construction of
the relaxed current operator is possible because
the environment is explicitly taken into account.

Because we are interested in the weak tunneling limit,
we take $H_{qp}+H_{env}$ as unperturbed hamiltonian and
$H_T$ as perturbation. Both $H_{qp}$ and $H_{env}$ are
quadratic and the perturbation calculation
is straightforward.

Current-voltage characteristic can be obtained from
the perturbation theory and
one obtains
\begin{equation}
I_T(V)=I_R(V)={2e \over \hbar^2}\mbox{ Im}\,\left(
  X_{ret}\left(-eV/\hbar \right) \right) \, ,
\end{equation}
where
\begin{equation}
X_{ret}(\omega)=-i\int_{-\infty}^{\infty}dt \, \Theta (t)
  e^{i\omega t}
  \left\langle \left[ A(t)e^{-i\tilde{\varphi}(t)},A^{\dagger}(0)
   e^{i\tilde{\varphi}(0)} \right] \right\rangle_0 \, ,
\end{equation}
and $A(t)=\sum_{kq\sigma}T_{kq}
c_{q\sigma}^{\dagger}(t)c_{k\sigma}(t)$.
$\langle \cdots \rangle_0$ represents the average over unperturbed
state and all operators are written in interaction picture.
Because the expectation values of the tunneling current and
the relaxed current are the same, we will drop the subscript.
By explicitly evaluating the average, one obtains
\begin{equation}
I(V)={1 \over eR_T}(1-e^{-\beta eV})\int_{-\infty}^{\infty} dE
  {E \over 1-e^{-\beta E}} P(eV-E) \, ,
\label{eqn:IV}
\end{equation}
where
\begin{equation}
P(E)={1 \over 2\pi \hbar} \int_{-\infty}^{\infty}dt \,
  \exp\left[\left\langle \left[\tilde{\varphi}(t)-
  \tilde{\varphi}(0)\right]
  \tilde{\varphi}(0) \right\rangle_0 +iEt/\hbar \right] \, .
\label{eqn:probability}
\end{equation}
Eq.(\ref{eqn:IV}) together with Eq.(\ref{eqn:probability})
agrees with the result of Ref.\cite{Devoret},
which justifies two current operators.

All information of the environment is stored in the phase-phase
correlation function
$J(t)=\langle [\tilde{\varphi}(t)-\tilde{\varphi}(0)]
  \tilde{\varphi}(0) \rangle_0$ and
from the fluctuation-dissipation theorem, $J(t)$ can be related
to the impedance $Z(\omega)$,
\begin{equation}
J(t)=2\int_{0}^{\infty} {d\omega \over \omega}
 {\mbox{Re } Z_t(\omega) \over R_Q}
 \left\{ \coth{\beta \hbar \omega \over 2}
 \left[\cos\omega t-1 \right] -i\sin\omega t \right\} \, ,
\label{eqn:correlation}
\end{equation}
where $Z_t(\omega)=\left( Z^{-1}(\omega)-i\omega C \right)^{-1}$
and $R_Q$ is the resistance quantum $h/e^2$.

Let us present current-voltage characteristics in a few
special cases. It illustrates the effect of the
environment and we later find that $S_{I_T}(\omega)$
is closely related to the characteristic.
Firstly, let us take $Z(\omega)=0$.
Then from Eq.(\ref{eqn:probability},
\ref{eqn:correlation}), one finds $P(E)=\delta(E)$ and
the junction shows Ohmic behavior.
The charging energy plays no role in
the low impedance environment. Secondly, let us take $Z(\omega)
=R\gg R_Q$ and $\beta E_C\gg 1$.
Then from Eq.(\ref{eqn:probability},\ref{eqn:correlation}),
one finds $P(E)=\delta(E-E_C)$ and the characteristic develops
a gap and current does not flow for $|eV|<E_C$.
This agrees with the prediction of
Likharev {\it et al}\cite{Likharev}
and one realizes that their semiclassical theory becomes valid
if a junction is connected to high impedance environment.
Finally, let us take Ohmic environment of arbitrary resistance,
$Z(\omega)=R$\cite{realistic}.
In this case, the characteristic is
rather complicated and we present only the limiting behaviors
at zero temperature.
With small voltage bias $|eV|\ll E_C$, one finds\cite{Devoret}
\begin{equation}
I(V)={\exp(-2\gamma R/R_Q) \over \Gamma(2+2R/R_Q)}
  {V \over R_T}\left[ {\pi R \over R_Q}{ |eV| \over E_C}
  \right]^{2R/R_Q} \, ,
\label{eqn:anomaly}
\end{equation}
where $\gamma=0.577\ldots$ is the Euler constant.
Note that Eq.(\ref{eqn:anomaly}) smoothly interpolates
the characteristics of the low and the high impedance environments.
Also note that with the finite impedance,
the suppression due to Coulomb blockade effect is
not strong enough to develop well defined gap even at zero
temperature.
With large voltage bias $eV \gg E_C$, one finds\cite{Ingold}
the characteristic approaches the high impedance result,
\begin{equation}
I(V)={1 \over eR_T}\left[eV-E_C+
  {R_Q \over \pi^2 R}{E_C^2 \over eV}\right] \, .
\end{equation}
The zero temperature current-voltage characteristics are
presented in Fig.\ref{fig:IV}.
For the experimental measurement of the current-voltage
characteristic, see Ref.\cite{Cleland}.

\section{Noise spectrum of tunneling current}
\label{sec:tunnelingnoise}
\subsection{Noise spectrum of tunneling current}
Noise spectral density of current or noise spectrum
is defined
as Fourier transform of a current-current correlation
function,
\begin{equation}
S_I (\omega)={1 \over \pi}\int dt \, e^{i\omega t}
  \left( {1 \over 2}\left\langle \left\{ \hat{I}(t), \hat{I}(0)
  \right\} \right\rangle - \langle \hat{I}(t)\rangle
  \langle \hat{I}(0)\rangle \right) \, .
\end{equation}
Because we later find the noise spectrum of the tunneling
current different from the one of the relaxed current,
we will put subscript $T$ and $R$ to give distinction.

Calculation of $S_{I_T}(\omega)$ is as straightforward as
the calculation of the current-voltage characteristic
and one finds
\begin{eqnarray}
S_{I_T}(\omega)&=& {1 \over 2\pi R_T}\int dE \,
  {E\over 1-e^{-\beta E}}
  \label{eqn:SITprelim} \\
  & & \times \sum_{\pm} P(\hbar\omega \pm eV-E)
  \left(1+e^{-\beta(\hbar\omega \pm eV)}\right) \, .
  \nonumber
\end{eqnarray}
Structure of Eq.(\ref{eqn:SITprelim}) is similar to
the formula for the current-voltage characteristic Eq.(\ref{eqn:IV})
and indeed, $S_{I_T}(\omega)$ can be written in terms of
$I(V)$,
\begin{equation}
S_{I_T}(\omega)={e \over 2\pi}\sum_{\pm}
  \coth{\beta(eV \pm \hbar\omega) \over 2}
  I\left(V \pm \hbar\omega/e\right)
  \, .
\label{eqn:SIT}
\end{equation}
Eq.(\ref{eqn:SIT}) allows one to obtain the noise spectrum
of the tunneling current from the current-voltage characteristic.

It is a simple task to verify that Eq.(\ref{eqn:SIT})
satisfies the fluctuation-dissipation theorem in equilibrium $V=0$.
To verify it, one introduces a coupling of the tunneling
current to its conjugate force $f_T(t)$,
\begin{equation}
H_{f_T}=-\hat{I}_T f_{T}(t) \, ,
\end{equation}
and calculates the response function
$\chi_{T}(\omega)=\langle \hat{I}_T(t) \rangle_{\omega}/f_{T}(\omega)$.
Then, the fluctuation-dissipation theorem relates the response
to the noise spectrum
\begin{equation}
S_{I_T}(\omega)=\hbar \coth {\beta \hbar \omega \over 2}
 \mbox{Im }\chi_T (\omega) \, .
\end{equation}

In the following, we consider equilibrium and excess noise
for Ohmic environment at zero temperature.
Finite temperature noise will be discussed only when
it allows a simple analytic expression.

\subsection{Low impedance environment}
Let us first study the low impedance limit $Z(\omega)=R\ll R_Q$,
where the junction becomes Ohmic. In equilibrium,
Eq.(\ref{eqn:SIT}) reduces to Johnson-Nyquist noise\cite{Callen}
(Fig.\ref{fig:SITequil})
\begin{equation}
S_{I_T}(\omega)={1 \over \pi R_T}\hbar\omega
  \coth {\beta\hbar\omega \over 2}  \, .
\label{eqn:lowequilnoise}
\end{equation}

Excess noise is the difference between equilibrium and
non-equilibrium noise
$S^{X}_{I_T}(\omega;V)=S_{I_T}(\omega;V)-S_{I_T}(\omega;V=0)$.
The superscript $X$ is employed to denote the excess noise.

At zero temperature, the excess noise is absent for
$|\hbar\omega| > |eV|$ and for $|\hbar\omega| < |eV|$,
it shows piecewise linear dependence on frequency
(Fig.\ref{fig:SITnonequil}(a)),
\begin{equation}
S^{X}_{I_T}(\omega;V)={|eV|-|\hbar\omega| \over \pi R_T}
\label{eqn:SITXlow}
\end{equation}
Eq.(\ref{eqn:lowequilnoise}) and Eq.(\ref{eqn:SITXlow})
agree with Eq.(\ref{eqn:noiseDahm})
where the charging effect was neglected.
The agreement with the previous study illustrates
that the charging effect does not show up in
the low impedance environment.

\subsection{High impedance environment}
With $Z(\omega)=R \gg R_Q$, Eq.(\ref{eqn:SIT}) reduces to
\begin{equation}
S_{I_T}(\omega)={1 \over \pi R_T}(1+e^{-\beta\hbar\omega})
  \int_{-\infty}^{\infty}dE \, {E \over 1-e^{-\beta E}}
  P(\hbar\omega-E) \, ,
\label{eqn:SITPE}
\end{equation}
in equilibrium where
\begin{equation}
P(E)= {1 \over \sqrt{4\pi E_C k_B T}}
  \exp\left[- { (E-E_C)^2 \over 4E_C k_B T} \right]  \, .
\label{eqn:P}
\end{equation}
At high temperature $k_B T \gg E_C$, $S_{I_T}(\omega)$ becomes
$(2k_B T /\pi R_T)(1-E_C/3k_B T+\mbox{$\cal O$}(E_C/k_B T)^2 )$
for $\hbar\omega \ll E_C$, which is the Coulomb blockade correction
to white noise. At low temperature $k_B T \ll E_C$,
$S_{I_T}(\omega)$ becomes $(2E_C /\pi R_T)\exp\left(
-E_C/k_B T\right)$ for $\hbar\omega \ll k_B T$.
The Coulomb blockade strongly suppresses the noise
far below the white noise level in the low temperature
(Fig.\ref{fig:SITequil}).

Both high and low temperature limiting behaviors agree with
earlier calculation of Ben-Jacob {\it et al}\cite{Ben-Jacob}
where they calculated the equilibrium noise spectrum of
an open junction which is not connected
to the external circuit.
Ben-Jacob {\it et al}'s result can be cast into
the form of Eq.(\ref{eqn:SITPE}) with $P(E)$ replaced by
\begin{equation}
P_{BJ}(E)=\left[\sum_{\mbox{integer}\, q}
  \exp\left(-{q^2 E_C \over k_B T}\right)\right]^{-1}
  \sum_{q}\delta\left(E-E_C(2q+1)\right)\exp\left(
  -{q^2 E_C \over k_B T} \right) \, .
\label{eqn:Pbj}
\end{equation}
Note that $P_{BJ}(E)$ is a discrete form of $P(E)$.
$P(E)$ and $P_{BJ}(E)$ become identical in
the high temperature limit $k_B T \gg E_C$
whereas they exhibit delicate
difference in the low temperature limit $k_B T \ll E_C$.
Recalling that $P(E)$ is a probability to emit energy $E$
to environment\cite{Ingold}, one finds that Eq.(\ref{eqn:Pbj})
implies the quantization of the emitted energy
whereas Eq.(\ref{eqn:P}) contains continuous spectrum of
the emitted energy.

The charge in each electrode is quantized
both in the open junction and in the closed junction
with high impedance environment. Hence the energy used to
charge the electrodes has only discrete values.
In the open junction, the whole emitted energy is
used to charge the electrodes.
In the junction closed by the environment, however,
the emitted energy is used only in part to charge
the electrodes and the rest excites the environmental
degrees of freedom.
Therefore $P(E)$ has continuous spectrum of energy
whereas $P_{BJ}(E)$ allows only discrete energy values.

In the high impedance environment, the excess noise
shows piecewise linear dependence on frequency
at zero temperature(Fig.\ref{fig:SITnonequil}(a,b)).
For small voltage bias $|eV|<E_C$, the excess noise
has finite spectral density(Fig.\ref{fig:SITnonequil}(a)),
\begin{equation}
S^{X}_{I_T}(\omega)={ \left|eV\right|-
  \left|\left|\hbar\omega\right|-E_C\right|
  \over 2\pi R_T} \, ,
\end{equation}
for $-|eV|<|\hbar\omega|-E_C<|eV|$.
Note that even though current does not flow when $|eV|<E_C$,
the excess noise does not vanish
around $\hbar\omega=\pm E_C$.
At higher voltage bias, finite current flows and
Fig.\ref{fig:SITnonequil}(b) shows the evolution
of the excess noise with the voltage.
The excess noise again shows the piecewise linear
dependence. The piecewise linear dependence in
both the low and the high impedance environments
originate from the narrow energy range of $P(E)$.

\subsection{Ohmic impedance environment}
One can also obtain $S_{I_T}(\omega)$ in arbitrary
Ohmic impedance environment $Z(\omega)=R$.
In the low frequency regime $|\hbar\omega|\ll E_C$,
the zero temperature equilibrium fluctuations are
suppressed by the Coulomb blockade effect,
\begin{equation}
S_{I_T}(\omega)={\exp(-2\gamma R/R_Q) \over \pi \Gamma(2+2R/R_Q)}
  {|\hbar\omega| \over R_T}\left[{\pi R \over R_Q}
  {|\hbar\omega| \over E_C}\right]^{2R/R_Q} \, .
\end{equation}
However
the suppression is not complete even at zero temperature.
The degree of the suppression depends on the impedance and
the higher the impedance, the stronger the suppression is.

In the high frequency regime $|\hbar\omega|\gg E_C$,
the equilibrium noise eventually
converges to the result of the high impedance environment,
\begin{equation}
S_{I_T}(\omega)={e \over \pi R_T}\left[
  |\hbar\omega|-E_C+{R_Q \over \pi^2 R}
  {E_C^2 \over |\hbar\omega|}
  \right]  \, ,
\end{equation}
and
the convergence is faster with the higher impedance
(Fig.\ref{fig:SITequil}).

With finite Ohmic impedance,
the excess noise no longer exhibits piecewise linear
dependence on the frequency because $P(E)$
allows broad range of energy.
Since the Coulomb blockade
effect is blurred by the quantum fluctuations of charge
in the finite impedance environment,
the excess noise is finite in broad frequency range
(Fig.\ref{fig:SITnonequil}(a)).
At zero frequency,
\begin{equation}
S^{X}_{I_T}(\omega=0)={1 \over \pi}
  {\exp(-2\gamma R/R_Q) \over \Gamma(2+2R/R_Q)}
  {|eV| \over R_T}\left[ {\pi R \over R_Q}{ |eV| \over E_C}
  \right]^{2R/R_Q} \, .
\end{equation}

\subsection{Noise power}
Before we close the discussion of the tunneling current
noise spectrum, let us mention one implication of
Eq.(\ref{eqn:SIT}).
In the presence of finite current, the correlation between
charge tunneling events can be estimated from the noise power $P$,
which is defined as $P=2\pi S_I(\omega=0)$.
If there is no correlation between the tunneling events,
the noise power retains its full value $P=2eI$.
If there is any kind of correlation which regulates tunneling,
it decrease the noise power below its full value.
For example, in a single channel quantum point contact,
the Fermi correlation is known to reduce the noise power to
$P=2eI(1-{\cal T})$ where $\cal T$ is the transmission coefficient
\cite{Lesovik,Yurke,Buttiker}.

{}From Eq.(\ref{eqn:SIT}), one finds the noise power
of the tunneling current to have the full value
independent of the environment,
\begin{equation}
P_T=2eI(V) \, ,
\end{equation}
where the subscript $T$ is employed to denote
that this is the noise power
of the tunneling current.
The full noise power implies that
Fermi statistics and the charging energy do not
give rise to any correlation between tunneling events.
This result is understandable since we are studying the weak
tunneling limit and the time interval between two consecutive
tunneling events is assumed to be large.

\section{Noise spectrum of relaxed current}
\label{sec:relaxednoise}
\subsection{Noise spectrum of relaxed current}
Most theoretical works on current fluctuations
calculates the noise spectrum from electron crossing
of a particular plane in a circuit. However,
as Landauer and Martin state\cite{Landauer}
`Currents are measured either through the voltage
developed across a small series resistor in the circuit,
or through the magnetic fields generated by the displacement
current. Contrary to occasional perceptions, noise measurements
have no relation to the crossing, by carriers, of a particular
plane along the circuit.'

To clarify the relation between conventional noise
measurements and noise spectrum treated in various
theoretical works
\cite{Lesovik,Yurke,Buttiker,Dahm,Ben-Jacob,Yang},
it is necessary to take an explicit account of
a measuring device in the formulation.
We assume that leads connected to a junction can
be used as a measuring device and calculate current
fluctuations at leads.
The relaxed current $\hat{I}_R$ corresponds to charge
flow at leads and therefore $S_{I_R}(\omega)$ corresponds
to spectral density of current fluctuations measured
at leads.
We will study the difference
between $S_{I_R}(\omega)$ and $S_{I_T}(\omega)$
in this section.

No tunneling limit $R_T \rightarrow \infty$ illustrates
a crucial difference of $S_{I_R}(\omega)$ and $S_{I_T}(\omega)$.
In this limit, the tunneling current operator
itself vanishes because it is proportional to the tunneling
matrix elements. Hence, one finds $S_{I_T}(\omega)=0$.
This result is natural since $\hat{I}_T$ is
the electron flow measured right at the tunneling barrier
and electrons cannot tunnel the barrier at all
in the no tunneling limit.
On the other hand, the relaxed current operator does not
depend on the tunneling matrix elements explicitly and
in the zero tunneling limit, one finds
$S_{I_R}(\omega)=S_{I_R}^{(0)}(\omega)$ where
\begin{equation}
S_{I_R}^{(0)}(\omega)={1 \over \pi}\mbox{Re }\left(
  {1 \over Z(\omega)-{1\over i\omega C}} \right)
  \hbar\omega\coth{\beta\hbar\omega \over 2} \, .
\label{eqn:SIR0}
\end{equation}
The superscript $(0)$ denotes that this is the zeroth
order contribution in the perturbation calculation.
Note that $S_{I_R}^{(0)}(\omega)$
is a generalization of Johnson-Nyquist
noise\cite{Callen}
$S_{I}(\omega)=1/(\pi R)\hbar\omega\coth(\beta\hbar\omega /2)$.
$S_{I_R}(\omega)$ measures the fluctuations of the relaxed
current in the environment
and $S_{I_R}(\omega)$ does not vanish even in
the no tunneling limit because of the zero point
fluctuations in the environment.

With finite tunneling, it is usually assumed that
$S_{I_R}(\omega)$ is a linear superposition of
the intrinsic noise $S_{I_R}^{(0)}(\omega)$ and
the tunneling noise $S_{I_T}(\omega)$, which
we call {\it linear superposition assumption}.
The assumption is frequently used to make comparison
of experiments with theoretical calculations
(for example, see Ref.\cite{Liefrink}).

Explicit calculation of $S_{I_R}(\omega)$
to the second order in the tunneling matrix
produces
\begin{eqnarray}
S_{I_R}(\omega)&=&S_{I_R}^{(0)}(\omega)+S_{I_R}^{(2)}(\omega)
  \, \label{eqn:SIR} \\
S_{I_R}^{(2)}(\omega) &=& S_{I_R}^{(2A)}(\omega)+
  S_{I_R}^{(2B)}(\omega) + S_{I_R}^{(2C)}(\omega)
  \nonumber \, ,
\end{eqnarray}
where
\begin{eqnarray}
S_{I_R}^{(2A)}(\omega) &=& {1 \over \left| 1-i\omega C Z(\omega)
  \right|^2 } S_{I_T}(\omega)
  \, , \label{eqn:SIR2} \\
S_{I_R}^{(2B)}(\omega) &=&  -{e \over \pi}
   \left\{ \mbox{Im }{1 \over 1-i\omega C Z(\omega)}
     \right\}^2 \coth{\beta\hbar\omega \over 2}
  \sum_{\pm} \pm I(V \pm \hbar\omega /e) \, \nonumber \\
S_{I_R}^{(2C)}(\omega) &=& {e^2 \over \pi\hbar^2}\mbox{Im }\left(
  {1 \over \left\{1-i\omega C Z(\omega) \right\}^2} \right)
  \coth{\beta\hbar\omega \over 2} \nonumber  \\
 & &  \times \mbox{Re }\sum_{\pm}\left\{
  X_{ret}(-eV/\hbar \pm \omega)-X_{ret}(-eV/\hbar)
   \right\} \, . \nonumber
\end{eqnarray}
The superscript $(2)$ in $S_{I_R}^{(2)}(\omega)$
denotes that this is the second order
contribution in the perturbation calculation.
Unlike the intrinsic noise, $S_{I_R}^{(2)}(\omega)$
depends on the voltage.

At zero frequency, one finds
$S_{I_R}^{(2A)}(0)=S_{I_T}(0)$ and
$S_{I_R}^{(2B)}(0)=S_{I_R}^{(2C)}(0)=0$, so that
that the linear superposition
assumption holds at zero frequency for arbitrary environment.
Because $S_{I_R}^{(0)}(\omega)$ vanishes at zero frequency,
one also finds that noise powers of the tunneling current
and the relaxed current are always the same
$P_R=P_T$.
For non-zero frequency, one finds that the assumption
holds if
\begin{equation}
\left|Z(\omega)\right| \ll {1 \over \left|i \omega C
  \right|} \, ,
\label{eqn:valid}
\end{equation}
and it does not otherwise.
Eq.(\ref{eqn:valid}) has interesting implications.
For example, in the Ohmic environment with
high impedance $Z(\omega)=R \gg R_Q$
where Coulomb blockade effect is clear,
the excess noise of the tunneling
current exhibits interesting feature at frequencies
around $\hbar\omega = \pm E_C$. However Eq.(\ref{eqn:valid})
predicts that at this frequency range,
the excess noise of the relaxed current
is noticeably different from the tunneling excess noise
since Eq.(\ref{eqn:valid}) translates to
$|\hbar\omega| \ll (R_Q/R)E_C$(see
Sec.\ref{subsec:high}).

Let us first discuss the equilibrium noise.
Due to the fluctuation-dissipation theorem,
the equilibrium noise spectrum can be alternatively
derived by coupling the relaxed current to its conjugate force
\begin{equation}
H_{f_R}=-\hat{I}_R f_R(t) \, ,
\end{equation}
and calculating the response function.
This procedure produces
\begin{eqnarray}
S_{I_R}(\omega)&=& S_{I_R}^{(0)}(\omega)
  -{e^2 \over \pi \hbar^2}\coth {\beta\hbar\omega \over 2}
  \mbox{Im }\left\{
  {1 \over \left( 1-i\omega C Z\left(\omega\right)\right)^2 }
  \right. \label{eqn:SIR2eq} \\
  & & \left. \times
   \left( X_{ret}(\omega)+X_{ret}^{*}(-\omega)+
   -X_{ret}(0)-X_{ret}^{*}(0) \right) \right\}
   \, . \nonumber
\end{eqnarray}
It can be verified that Eq.(\ref{eqn:SIR2eq})
agrees with Eq.(\ref{eqn:SIR}) in equilibrium.
Note that though both $S_{I_R}(\omega)$ and $S_{I_T}(\omega)$
satisfy the fluctuation-dissipation theorem,
they are still different since
they are related to different response functions.

\subsection{Low impedance environment}
\label{subsec:low}
To make predictions from Eq.(\ref{eqn:SIR}),
one has to evaluate the real part of the retarded
Green's function $X_{ret}(\omega)$.
For uniform tunneling density of states,
one finds at zero temperature
\begin{eqnarray}
& &\mbox{Re }\sum_{\pm}\left\{ X_{ret}(-eV/\hbar\pm \omega)
  -X_{ret}(-eV/\hbar) \right\}
  \\
&=& {\hbar^2 \over 2\pi e^2 R_T}\int dE\, (P(E)-P(-E))
   \nonumber \\
& & \times \sum_{\pm}\left\{
  (eV-E \pm \hbar\omega)\ln\left|eV-E \pm \hbar\omega\right|
  -(eV-E)\ln\left|eV-E\right| \right\}
  \, . \nonumber
\end{eqnarray}

In the low impedance environment $Z(\omega)=R\ll R_Q$,
one finds $S^{(2C)}_{I_R}(\omega)$ vanishes at
zero temperature and the zero temperature equilibrium
noise becomes
\begin{equation}
S_{I_R}(\omega)=S_{I_R}^{(0)}(\omega)+{1 \over \pi R_T}
  \mbox{Re }\left( {1 \over \left( 1- i\omega C Z(\omega)
  \right)^2 } \right) |\hbar \omega| \, .
\label{eqn:lowequilR}
\end{equation}
It can be verified that Eq.(\ref{eqn:lowequilR})
agrees, to the leading order in $1/R_T$,
with zero temperature equilibrium noise in
the presence of a shunt resistor $R_T$ in parallel
with a capacitor $C$
\begin{equation}
S_I(\omega)={1 \over \pi}\mbox{Re }\left(
 {1 \over Z(\omega)+{1 \over -i\omega C +1/R_T}}
 \right) |\hbar \omega| \, .
\end{equation}

When the voltage bias is applied, one finds the zero
temperature excess noise
\begin{equation}
S^{X}_{I_R}(\omega)={1 \over 1+\omega^2 R^2 C^2}
  S^{X}_{I_T}(\omega) \, .
\label{eqn:suppression}
\end{equation}
Eq.(\ref{eqn:suppression}) can be understood if
we assume that current fluctuations induced by
tunneling is incoherent with equilibrium zero
point current.

Then the behavior of a junction is essentially classical.
The only effect of the tunneling is to introduce
additional charge to plates of capacitor and
from a conventional circuit analysis of a capacitor
connected to impedance $Z(\omega)$\cite{commentshunt},
one finds the current induced by tunneling is\cite{Ingold}
\begin{equation}
I_R(t)=\int ds \, {\cal R}(t-s)I_T(s) \, ,
\end{equation}
where ${\cal R}(t)=0$ for $t<0$ and
${\cal R}(\omega)=1/(1-i\omega C Z(\omega))$.
{}From the incoherence assumption, one immediately
finds
\begin{equation}
S^{X}_{I_R}(\omega)={1 \over |1-i\omega C Z(\omega)|^2}
  S^{X}_{I_T}(\omega) \, ,
\label{eqn:excessrelaxed}
\end{equation}
which agrees with Eq.(\ref{eqn:suppression}).

The factor $1/|1-i\omega C Z(\omega)|^2$ in
Eq.(\ref{eqn:excessrelaxed}) originates from the finite
relaxation time of the environment; the relaxation
of tunneled charge to the environment is not immediate
and it takes characteristic time of $CZ(\omega)$.

Fig.\ref{fig:excess}(a) shows the relaxed excess noise
in comparison to the tunneling excess noise.
At zero temperature,
there are three relevant energy scales $eV, \hbar/RC,
E_C$.
Fig.\ref{fig:excess}(a) assumes $eV \gg \hbar/RC \gg E_C$.
Due to the low impedance of the environment, the junction
shows Ohmic behavior. The tunneling excess noise(dashed line)
decays linearly with the frequency and its peak width
is $eV$. The relaxed excess noise(solid line) has
the same magnitude with the tunneling excess noise
at zero frequency. However, its peak width is given
as min$(eV,\hbar/RC)$ and for the case considered
in Fig.\ref{fig:excess}(a), it is $\hbar/RC$,
making the decay of the relaxed excess noise faster
than the tunneling excess noise.

\subsection{High impedance environment}
\label{subsec:high}
In the high impedance environment, the junction becomes
highly nonlinear and the shunt resistor model cannot
explain its behavior. To understand the junction behavior
in the high impedance environment, the investigation
of the zero point fluctuations in the environment
is needed.

In the no tunneling limit, the environment is decoupled
from quasi-particle degrees of freedom. The environmental
hamiltonian is quadratic and therefore it can be
diagonalized as a set of independent harmonic oscillators.
Each harmonic oscillator shows zero point fluctuations
of the magnitude $\langle x^2 \rangle= \hbar/(2m_0 \omega_0)
\coth\beta\hbar\omega_0 /2$ where $m_0$ and $\omega_0$
are mass and frequency of the oscillator, respectively.
Harmonic oscillator with frequency $\omega_0$ contribute
to the spectral density of the relaxed current at
$\omega=\omega_0$ and the zeroth order spectral
density $S^{(0)}_{I_R}(\omega)$ is a sum of such
contributions.

The situation becomes complicated if one turn on
tunneling. The tunneling couples each harmonic
oscillator to quasi-particle degrees of freedom and
other harmonic oscillators. Therefore the tunneling
provides each harmonic oscillator with the coupling
to an effective heat reservoir which consists of
other harmonic oscillators and quasi-particle degrees
of freedom.

The effect of such coupling
is twofold. It provides a noise source and it also
introduces dissipation. The noise source tends to
increase fluctuations in, for example, harmonic
oscillator displacement variables while the dissipation
has the opposite effect. It is known (for example, see
Ref.\cite{Chakravarty}) that the coupling to the reservoir
can even reduce the zero point point fluctuations
in certain situations.

The relaxed current $\hat{I}_R$ is linear in the
environmental variables. Therefore it can be written
as a linear combination of harmonic oscillator
displacements and the fluctuations of $\hat{I}_R$
is directly related to the fluctuations of
harmonic oscillators.

A harmonic oscillator coupled to the reservoir is
a difficult problem in general. However in the weak
tunneling limit, one can perturbatively calculate
the effect of the reservoir.

It is worth mentioning that the voltage bias is
a variable describing quasi-particle degrees of freedom.
Therefore fluctuations of harmonic oscillators
and the relaxed current depend on the voltage bias.

Fig.\ref{fig:excess}(b) shows the relaxed excess noise
in comparison to the tunneling excess noise
at zero temperature in the high impedance Ohmic
environment $Z(\omega)=R\gg R_Q$.
The relaxed excess noise can be interpreted as
change of harmonic oscillator fluctuations by
the voltage bias.
It is assumed that
$E_C>eV \gg \hbar/RC$. Note that current does not
flow because the voltage bias is below the gap
$eV<E_C$.
The tunneling excess noise shows two peaks at
$\hbar\omega = \pm E_C$.
In comparison, the relaxed excess noise(solid line)
becomes {\it negative} at $\hbar\omega \approx \pm E_C$.
The relaxed excess noise is multiplied by $10^4$ to
magnify the negative dips. Though the dips are quite
shallow, it is important to notice that the dips
imply the decrease of the total noise upon the
application of the voltage bias.
The dash-dotted line shows $1/(1+\omega^2 R^2 C^2)
S_{I_R}^{X}(\omega)$. The dash-dotted line is also
multiplied by $10^4$.
At higher voltage bias $eV>E_C$, the relaxed excess
noise shows a peak at $\hbar\omega=0$ and
shallow dips at $\hbar\omega \approx \pm E_C$.

{}From Fig.\ref{fig:excess}(b) and the relation
$S^{X}_{I_R}(0)=S^{X}_{I_T}(0)$, one finds that
the dissipative effect is dominant at
$\hbar\omega \approx \pm E_C$ while the noise source
effect is dominant at $\hbar \omega \approx 0$.

\subsection{Ohmic impedance environment}
Fig.\ref{fig:excess}(c) shows the two excess noises at
zero temperature in the Ohmic environment $Z(\omega)=R$.
It is assumed that $R=R_Q$ and $eV=0.5E_C$.
The tunneling excess noise(dashed line) shows two
peaks, while the relaxed excess noise(solid line)
shows a single peak at $\hbar\omega=0$.
The dash-dotted line shows $1/(1+\omega^2 R^2 C^2)
S_{I_T}^{X}(\omega)$ for comparison.
The dash-dotted line exhibit qualitatively
the same feature as $S_{I_R}(\omega)$ though
there are quantitative difference.

\subsection{Experimental implications}
Eq.(\ref{eqn:SIR}) makes one interesting prediction
in the low impedance Ohmic environment.
As illustrated in Sec.\ref{subsec:low}
(Fig.\ref{fig:excess}(a)), the excess noise measured from
the environment is different from the tunneling excess
noise and follows Eq.(\ref{eqn:suppression}).

Recently, there was an experiment which measured
the excess noise of the quantum point contact
in the low impedance environment\cite{Reznikov}.
The voltage bias $V$ of the order of $1$mV was applied
and the excess noise was measured at frequency
$\omega \sim 10$GHz which is much lower than
the voltage $\hbar\omega \ll eV$.
The frequency was chosen to avoid complications due to
other noise sources at low frequency such as $1/f$ noise.
The measured excess noise was compared with
theoretical calculations
of the noise power(for example, Ref.\cite{Yang} or
$S_{I_T}^{X}(\omega)$)
based on the assumption that the non-zero frequency
excess noise is essentially the same as the noise power
if $\hbar\omega \ll eV$.
The experiment showed nice agreement
with existing theories in ballistic regime.
However in the pinched off regime which corresponds to
the weak tunneling limit, it was found that
the noise was reduced to about $1/3$ of
theoretical predictions.

The quantum point contact in Ref.\cite{Reznikov}
is induced electrostatistically in the plane
of a 2DEG embedded in GaAs-AlGaAs heterostructure.
Due to the very different density of charge carriers
in metals and semiconductors, our calculation
on metallic junctions cannot be directly applied
to analyze the experiment. However we believe
that the difference of $S^{X}_{I_R}(\omega)$ and
$S^{X}_{I_T}(\omega)$ in the low impedance environment
is a generic property of the weak tunneling limit
because the difference originates from the finite
relaxation time of the environment which does not
depend on charge carrier density.

We argue that the noise suppression observed in
Ref.\cite{Reznikov} may have its origin
in the different frequency dependence of
the relaxed and tunneling excess noise
as illustrated in Sec.\ref{subsec:low}.
However, we are unable to establish quantitative
agreement of the experiment with Eq.(\ref{eqn:suppression})
because the precise values of $R$ and $C$ are not
available.
More experiment on, for example, the frequency dependence
of the suppression factor, is needed.

Eq.(\ref{eqn:SIR}) makes another interesting prediction
in the high impedance Ohmic environment.
At frequency $\hbar\omega \approx \pm E_C$, the spectral
density $S_{I_R}(\omega)$ measured at
the leads {\it decreases} below its equilibrium value
when the voltage bias is applied(Sec.\ref{subsec:high}),
making the relaxed excess noise negative.
Usually the spectral density increases when the voltage
bias is applied. Therefore the observation
of the negative excess noise can serve as
a crucial test of the present calculation.

\subsection{Weak scattering limit}
Calculation in this paper is restricted to the weak
tunneling limit $R_T \gg R_Q$.
It will be interesting to carry out similar study
for the weak scattering limit.
According to a recent experiment
on quantum point contact\cite{Reznikov},
the measured excess noises show the deviation from
existing theories only in the pinched off regime and
as a channel opens(as a transmission coefficient
$\cal T$ approaches $1$),
the deviation disappears.
If the deviation is due to the difference between
the relaxed excess noise and the tunneling excess noise,
the experiment implies that the junction in the weak
scattering limit no longer behaves as a conventional capacitor
and the difference between the two excess noise vanishes.
Though the experiment used junctions in semiconducting
heterostructure, we expect metallic junctions
show similar behavior at least in the low impedance
environment.

Also there was a theoretical study of a single
tunnel junction using the functional integral
technique\cite{Brown}.
It was claimed that the junction becomes purely
Ohmic and the capacitive effect vanishes
in the weak scattering limit.

In the weak scattering limit, electron transport
is continuous rather than being sporadic
as in the weak tunneling limit and
we don't expect a conventional capacitor is a proper
description of the junction.
However, further study is required to
gain the quantitative understanding
in the weak scattering limit.

\section{Conclusion}
In summary, we considered current transport through
a junction connected to electromagnetic
environment in the weak tunneling limit.
The charging energy is taken into account to
incorporate the electron-electron interaction effect.
We calculated both
the tunneling noise and the relaxed noise for arbitrary
environment and showed their dependence on
the charging energy and environment.
We found that contrary to conventional expectation,
the current through the potential barrier of the junction
and the current measured from the environment have
different excess noise.
In the low impedance environment, the difference is
due to the finite relaxation time of the junction.
In the high impedance environment, additional complication
occurs and at frequency $\hbar \omega \approx \pm E_C$,
the spectral density of the relaxed current decreases
below the equilibrium level when small voltage bias
is applied.

\section*{Acknowledgments}
We are grateful to M. Reznikov for kindly sending us
their manuscript before publication.
Research of L.L is partly supported by
Alfred Sloan Fellowship.

\begin{figure}
\caption{(a) Schematic diagram of the circuit.
  The circuit contains a single tunnel junction , voltage
  source, and the impedance $Z(\omega)$ of the environment.
  (b) The tunneling current and the relaxed current.
  The tunneling current represents the electron flow right
  at the potential barrier of a junction. As an example,
  the tunneling current through the insulating barrier
  in metal-insulator-metal junctions is drawn.
  The relaxed current represents the electron
  flow through the environment which is connected to
  a junction in series. In the figure, Ohmic environment
  with resistance $R$ is drawn.
}
\label{fig:circuit}
\end{figure}

\begin{figure}
\caption{Zero temperature current-voltage characteristics
  with $Z(\omega)=0.01R_Q$(solid line),
  $Z(\omega)=100R_Q$(dashed line), and
  $Z(\omega)=R_Q$(dash-dotted line). }
\label{fig:IV}
\end{figure}

\begin{figure}
\caption{Zero temperature equilibrium noise spectrum
  of the tunneling current for $Z(\omega)=0.01R_Q$(solid line),
  $Z(\omega)=100R_Q$(dashed line), and
  $Z(\omega)=R_Q$(dash-dotted line).
}
\label{fig:SITequil}
\end{figure}

\begin{figure}
\caption{(a) Zero temperature excess noise spectra of
  the tunneling current for $Z(\omega)=0.01R_Q$(solid line),
  $Z(\omega)=100R_Q$(dashed line) and
  $Z(\omega)=R_Q$(dash-dotted line). Voltage bias of
  $eV=0.5E_C$ is assumed.
  (b) Evolution of zero temperature tunneling
  excess noise with voltage bias in the high impedance
  environment $Z(\omega)=100R_Q$. $eV=0.5E_C$(solid line),
  $eV=1.5E_C$(dashed line), and $eV=2.5E_C$(dash-dotted line).
}
\label{fig:SITnonequil}
\end{figure}

\begin{figure}
\caption{
The excess noise spectrum of the relaxed current
$S_{I_R}^{X}(\omega)$(solid line) vs. the excess noise
spectrum of the tunneling current $S_{I_T}^{X}(\omega)$
(dashed line) at zero temperature. The dash-dotted line
shows $1/(1+\omega^2 R^2 C^2)S_{I_T}^{X}(\omega)$
for comparison.
(a) $Z(\omega)=0.01R_Q,\, eV=200E_C$. The dash-dotted line
  is not visible because it overlaps with the solid line.
(b) $Z(\omega)=100R_Q,\, eV=0.5E_C$. The solid line and
  the dash-dotted line are multiplied by $10^4$ to
  magnify their features. Note that $S_{I_R}^{X}(\omega)$
  is negative.
(c) $Z(\omega)=R_Q,\, eV=0.5E_C$. $S_{I_R}^{X}(\omega)$ has
  a single peak at $\omega=0$ compared to two peaks of
  $S_{I_T}^{X}(\omega)$. $S_{I_R}^{X}(\omega)$ deviates
  from $1/(1+\omega^2 R^2 C^2)S_{I_T}^{X}(\omega)$
  due to nontrivial behavior of the zero point
  fluctuations.
}
\label{fig:excess}
\end{figure}



\begin{references}
\bibitem{realistic} In real experiments,
  $Z(\omega)$ has frequency dependence
  and usually it decays at high frequency regime.
  However, we take Ohmic environment for illustration
  because Ohmic environment
  is simpler to treat and illustrates essential features of
  real experiments.
\bibitem{commentshunt} The shunt resistor gives only
  higher order effect and it can be ignored safely
  in the leading order calculation of the excess noise.
\end{references}
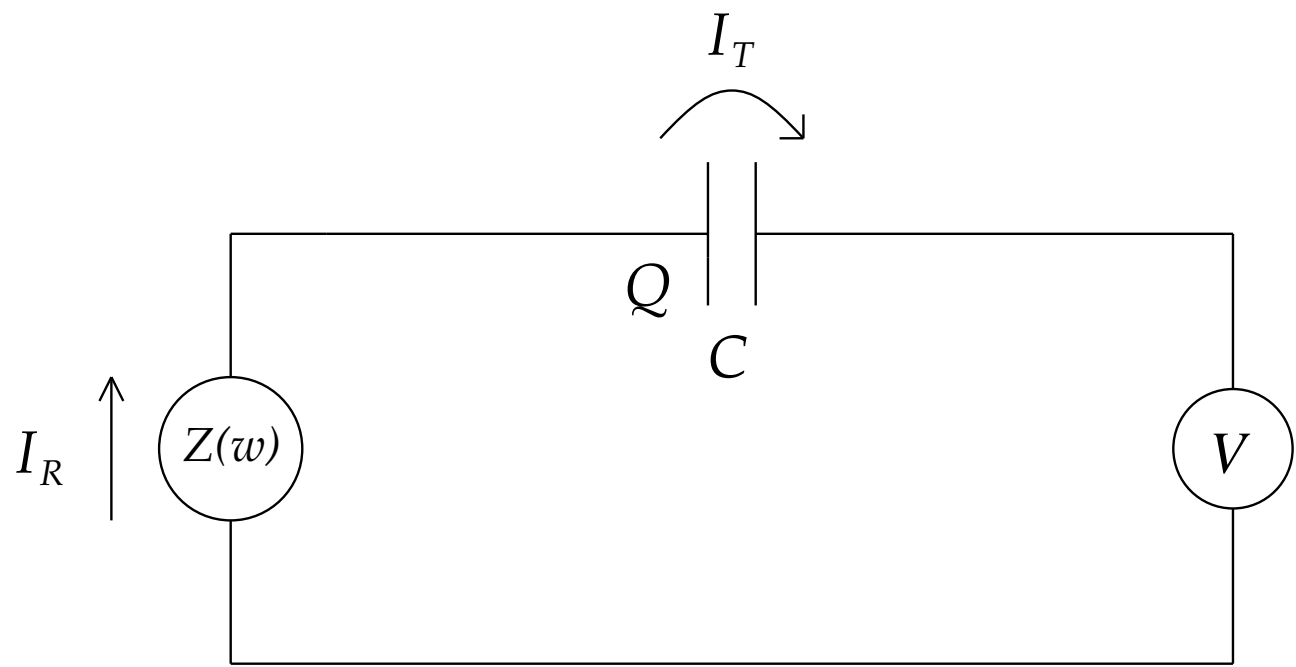
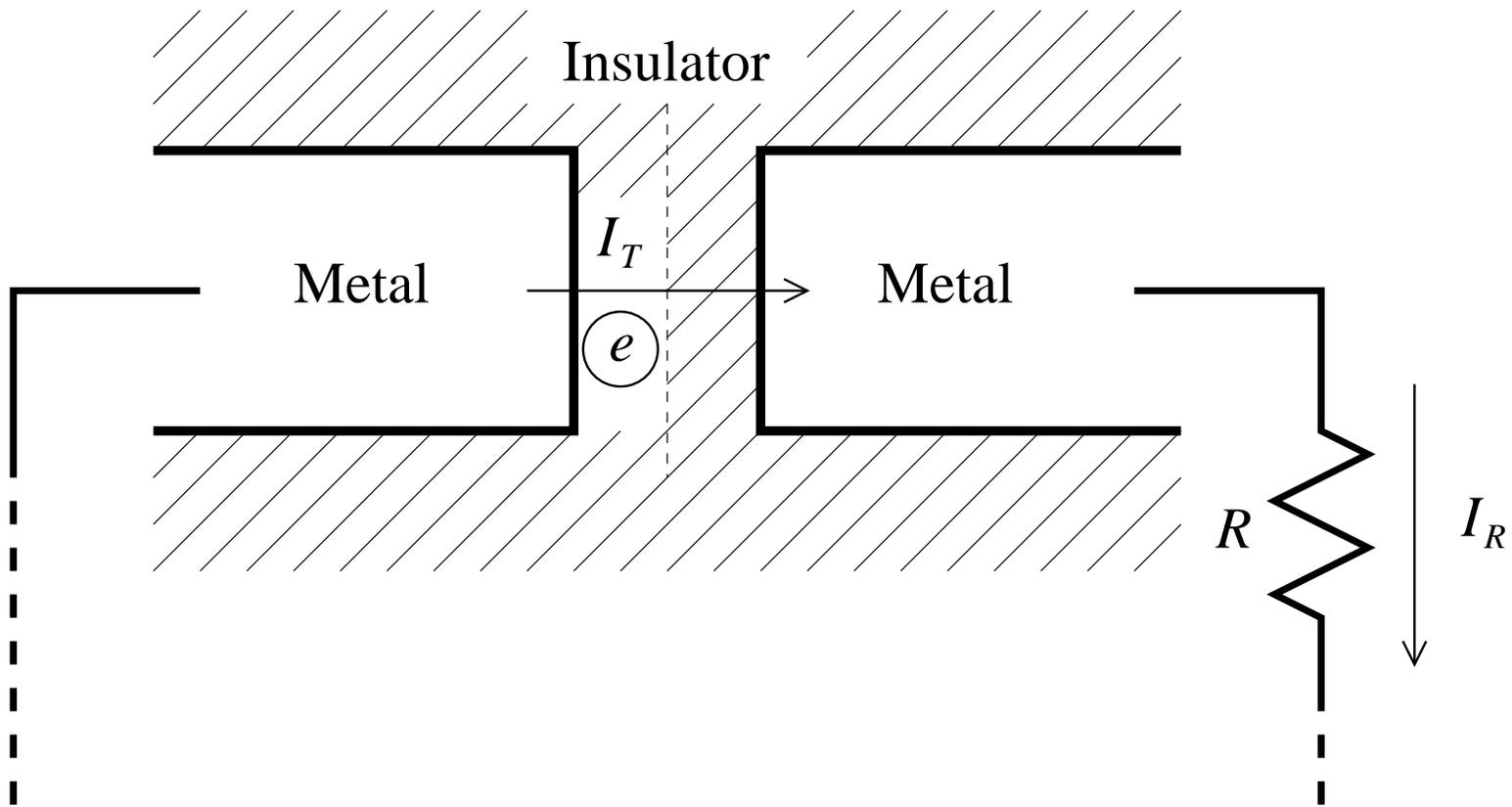
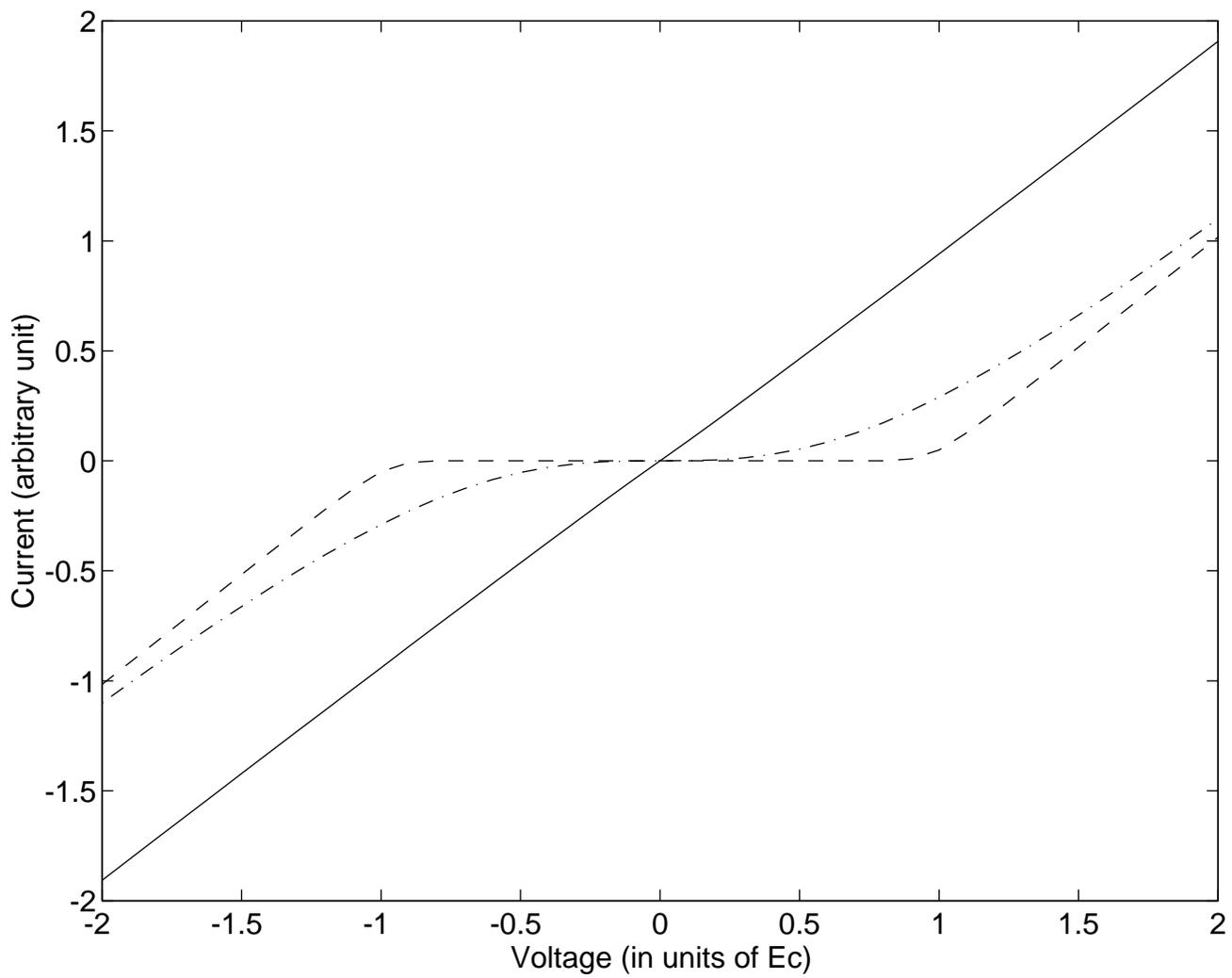
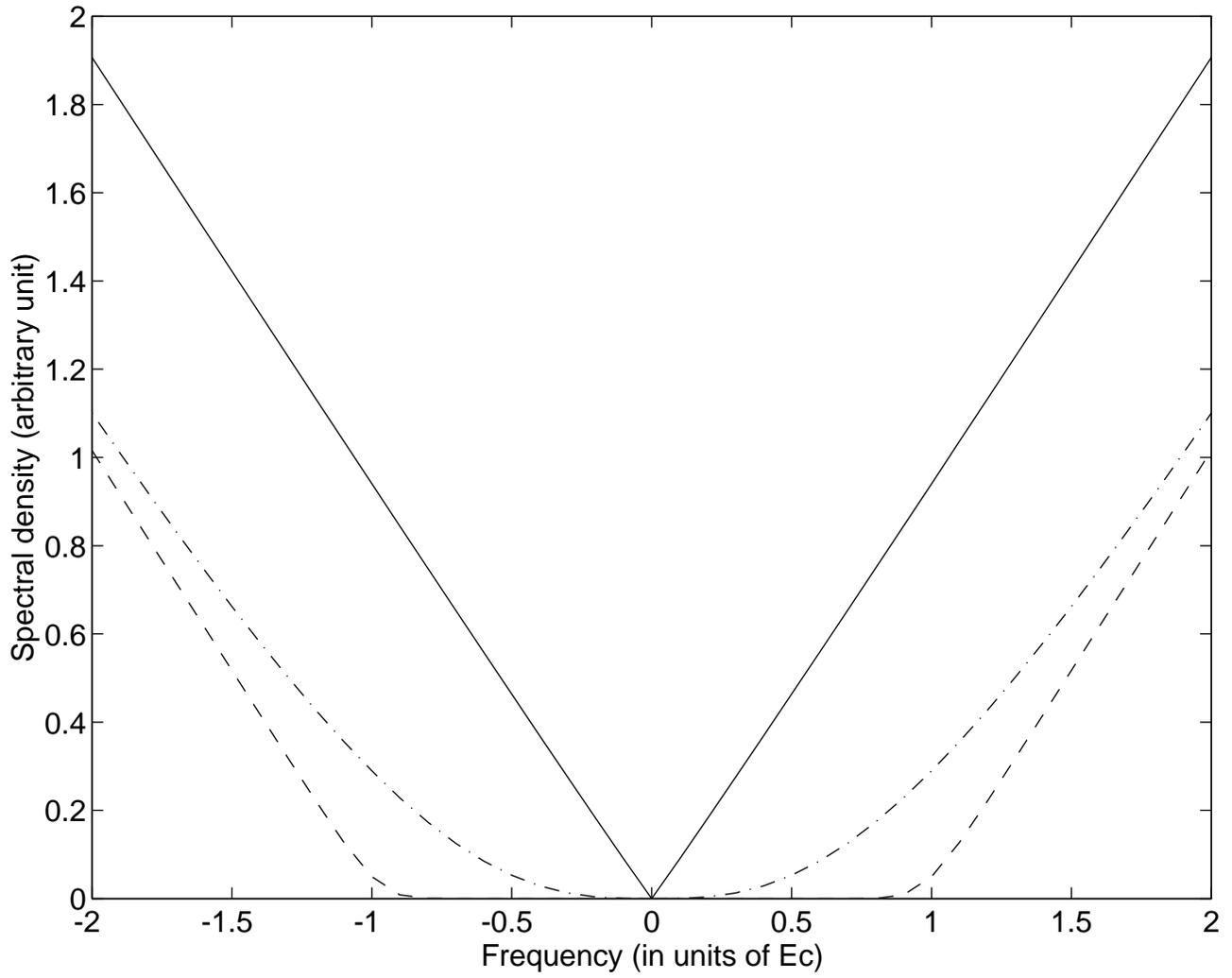
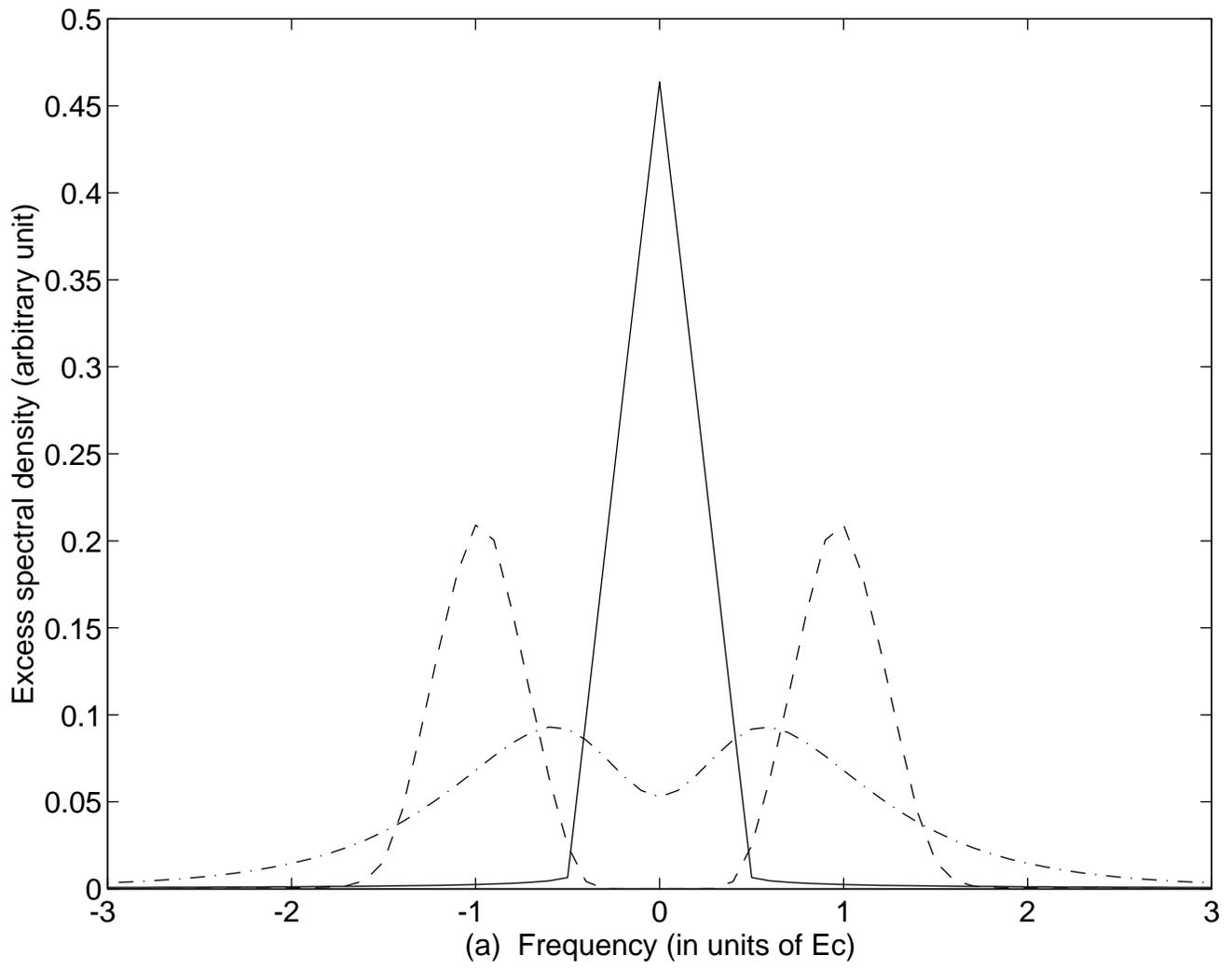
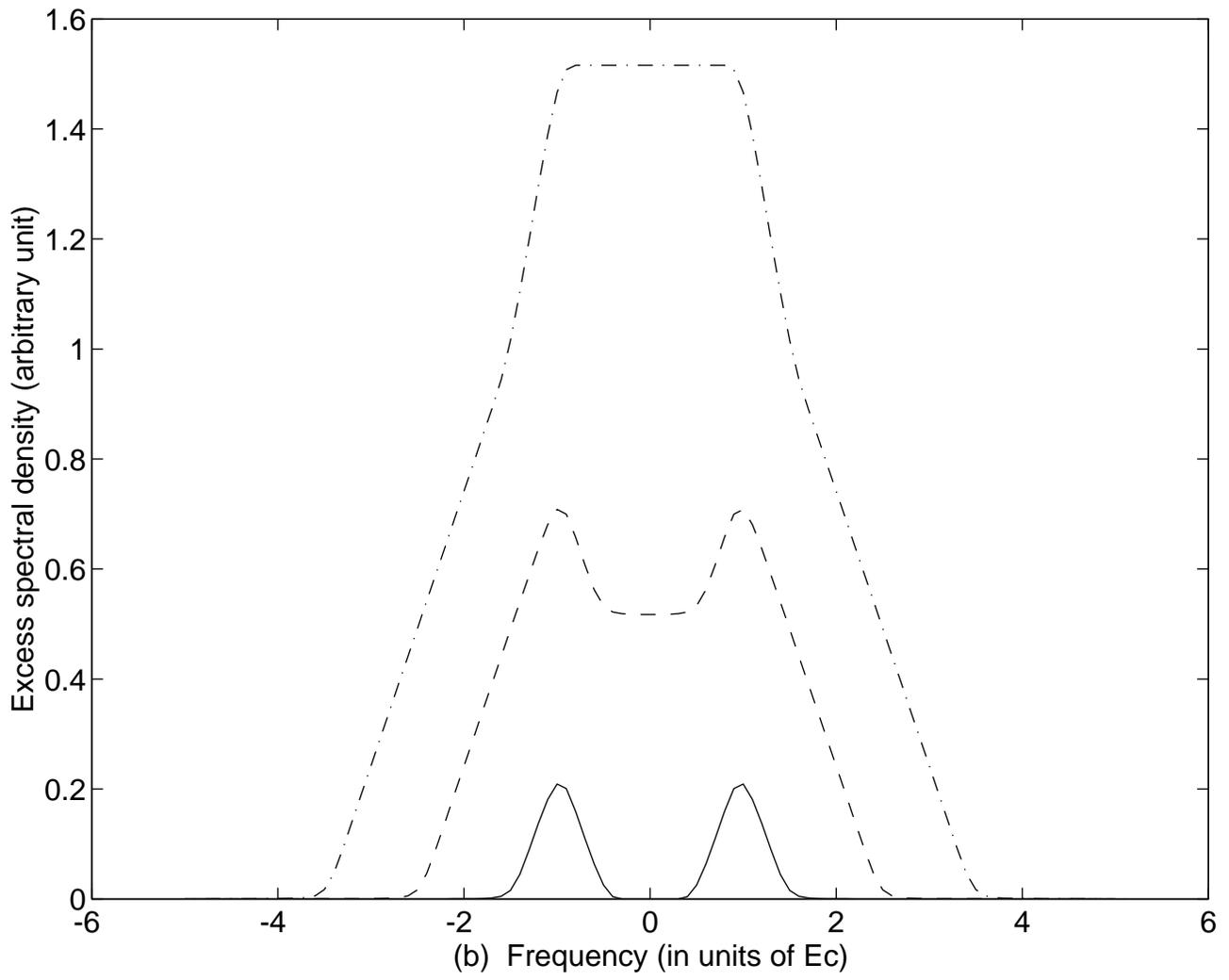
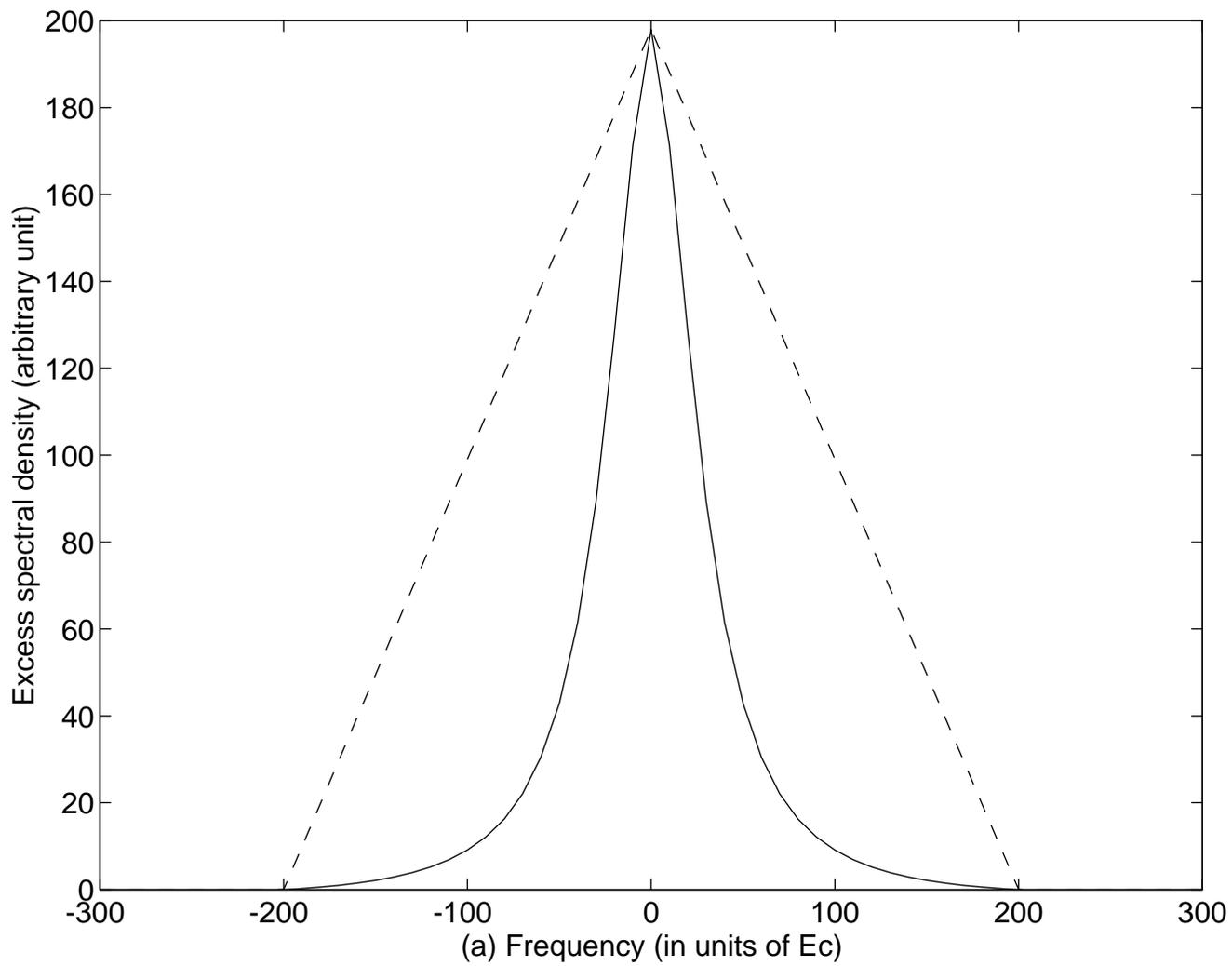
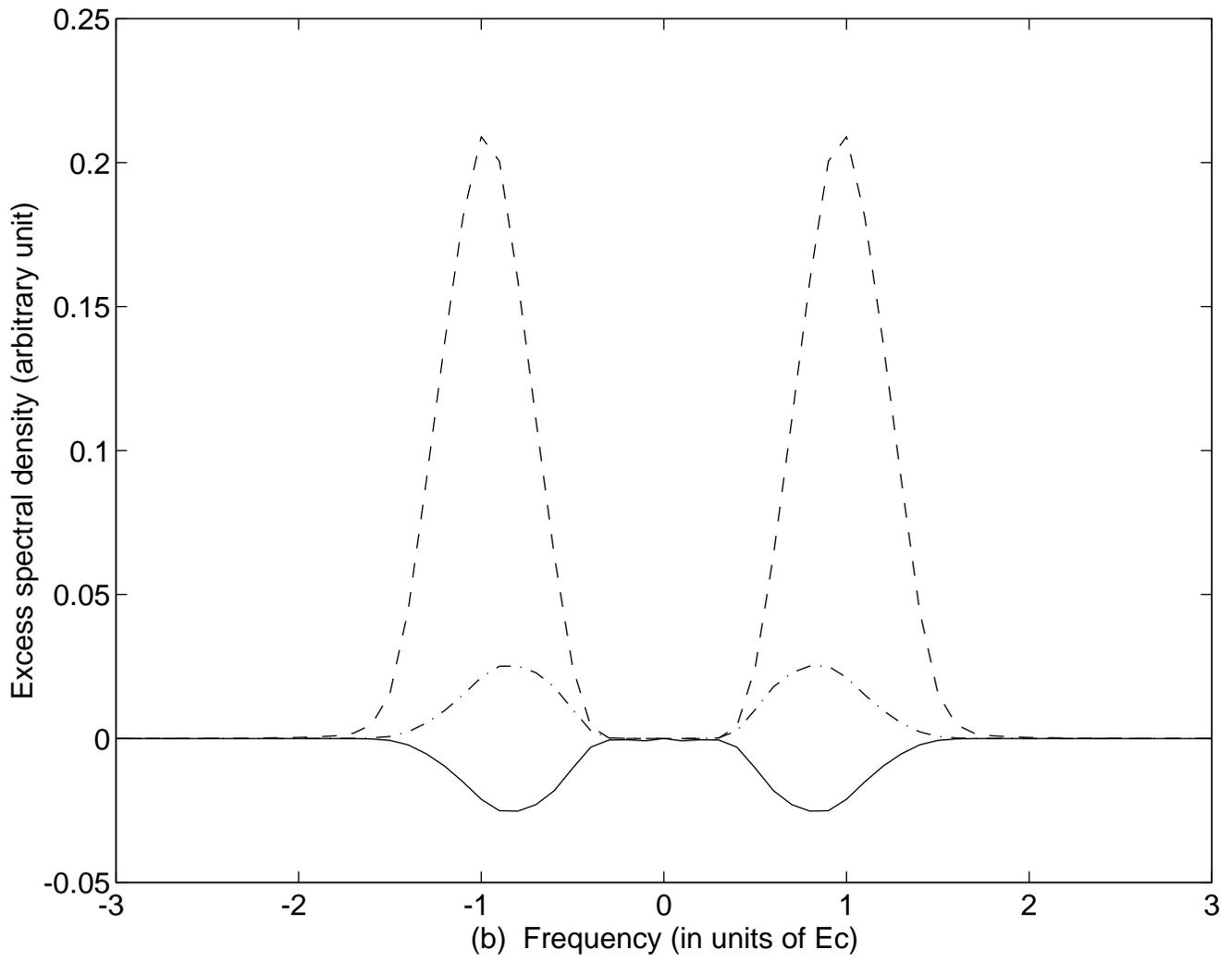
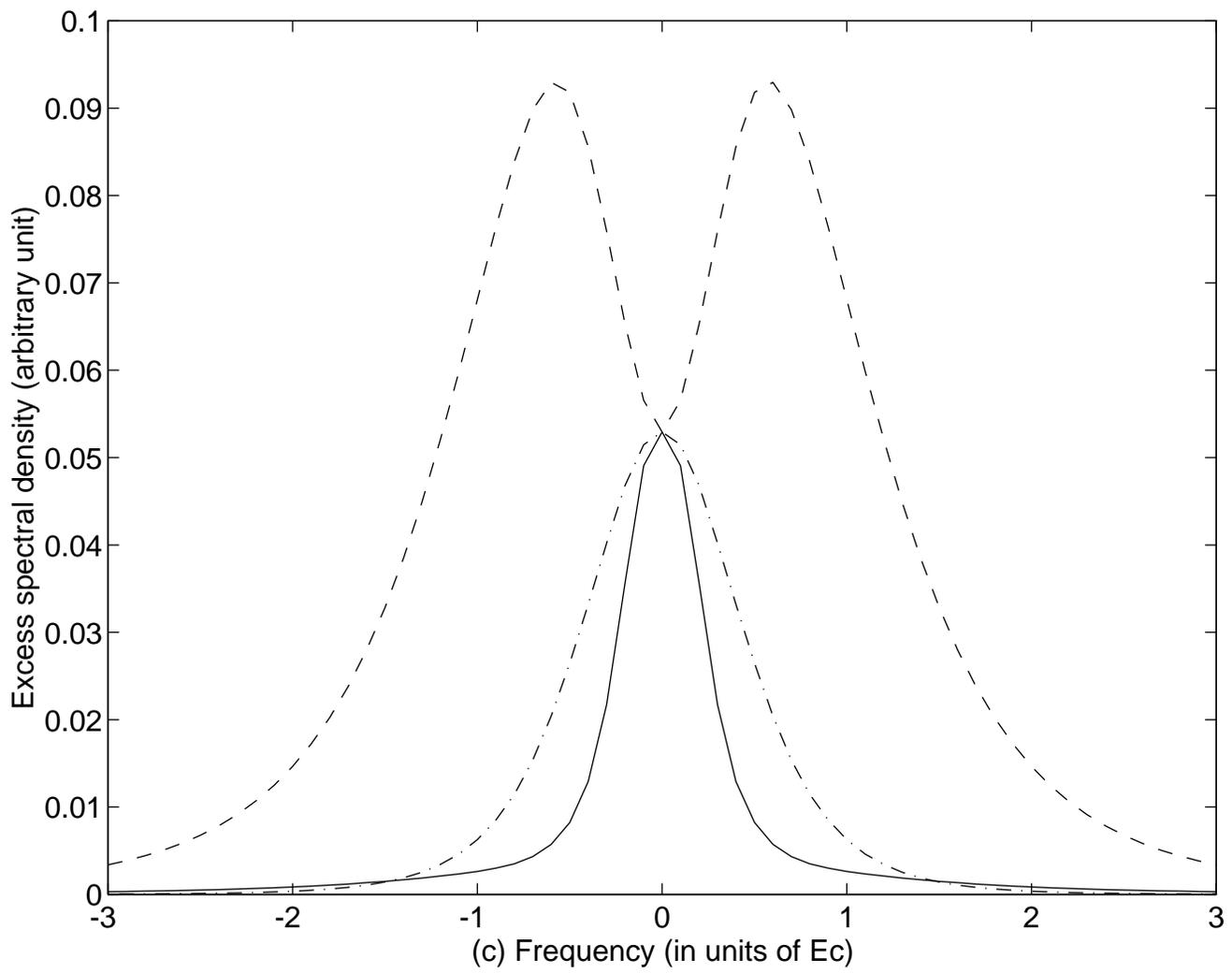
\end{document}